%% file: Manuscript.tex
\documentclass[preprint,11pt, 3p, sort&compress]{elsarticle}
\usepackage{fullpage}
\usepackage{color}
\usepackage{subcaption}
\usepackage{multirow}
\usepackage{amsmath}
\usepackage{amssymb}
\usepackage{hyperref}
\usepackage{bm}
\usepackage{pict2e}
\usepackage{epstopdf}
\usepackage{epsfig}
\usepackage{comment}
\usepackage{url}
\usepackage[ruled,vlined]{algorithm2e}

\newcommand{\normp}[1]{\| #1 \|}
\newcommand{\br}{\boldsymbol{\rho}}

\journal{Chemical Physics Letters}

\input{preamble}

\begin{document}

\begin{frontmatter}

\title{Periodic Pulay method for robust and efficient convergence acceleration of self-consistent field iterations}

\author[lbl]{Amartya S.\ Banerjee}
\ead{baner041@umn.edu}
\author[phanish_address]{Phanish Suryanarayana\corref{cor1}}
\ead{phanish.suryanarayana@ce.gatech.edu}
\author[pask_address]{John E.\ Pask}
\ead{pask1@llnl.gov}

\cortext[cor1]{Corresponding author}

\address[lbl]{Computational Research Division,
Lawrence Berkeley National Laboratory, Berkeley, CA 94720, U.S.A}
\address[phanish_address]{College of Engineering, Georgia Institute of Technology, Atlanta, GA 30332, U.S.A}
\address[pask_address]{Physics Division, Lawrence Livermore National Laboratory, Livermore, CA 94550, U.S.A}
\begin{abstract}
Pulay's Direct Inversion in the Iterative Subspace (DIIS) method is one of the most widely used mixing schemes for accelerating the self-consistent solution of electronic structure problems. In this work, we propose a simple generalization of DIIS in which Pulay extrapolation is performed at periodic intervals rather than on every self-consistent field iteration, and linear mixing is performed on all other iterations. We demonstrate through numerical tests on a wide variety of materials systems in the framework of density functional theory that the proposed generalization of Pulay's method significantly improves its robustness and efficiency. 
\end{abstract}


\end{frontmatter}

\section{Introduction} \label{sec:introduction}
Nonlinear equations are often posed as fixed point problems that lend themselves to solution via self-consistent iterations \citep{hoffman2001numerical, quarteroni2010numerical}. This is the commonly adopted practice in electronic structure calculations such as those based on density functional theory (DFT) \citep{hohenberg1964inhomogeneous, kohn1965self}, where additionally so-called \textit{mixing schemes} are routinely employed to accelerate convergence \citep{Martin_ES,Kohanoff}. The simplest of all mixing schemes is \textit{linear mixing}, which is an under-relaxed fixed-point iteration. Although linear mixing can be guaranteed to converge for many systems with the choice of small enough mixing parameter \citep{lin2013elliptic}, it tends to perform rather poorly in practice. Since the computational cost of electronic structure calculations is directly proportional to the number of self-consistent field (SCF) iterations required, considerable effort has been devoted to the formulation of more effective mixing schemes over the years, see, e.g., \cite{Martin_ES,Kohanoff} and references therein.  

Perhaps the most widely used mixing scheme is Pulay's Direct Inversion in the Iterative Subspace (DIIS) \citep{pulay_mixing, pulay_improved_scf}, based on Anderson's extrapolation \cite{Anderson1965}. Pulay's technique represents a specific variant of Broyden's quasi-Newton approach \cite{broyden1965class,bendt1982new,eyert1996comparative} and falls into the broad category of multisecant methods \citep{fang2009two}. Although the relative simplicity and overall performance of DIIS \citep{kudin2007converging} make it an attractive choice, it has been observed that Pulay mixing can stagnate and/or otherwise perform poorly in calculations involving certain metallic and/or inhomogeneous systems \citep{lin2013elliptic,Anderson_Accelaration_Convergence}. This has motivated the development of a number of alternative approaches, including variants of Broyden's method \cite{srivastava1984broyden,vanderbilt1984total,eyert1996comparative,marks2008robust}, the Relaxed Constrained Algorithm (RCA) \cite{cances2000can,cances2000convergence}, and a variety of preconditioning techniques \cite{lin2013elliptic,kerker1981efficient,ho1982dielectric,raczkowski2001thomas,anglade2008preconditioning}. However, while the improvements demonstrated have been in some cases substantial, increased complexity, additional parameters, and/or lack of transferability have hindered adoption in practice.

In this work, we introduce a simple generalization of the DIIS method for accelerating self-consistent field iterations, which we refer to as the Periodic Pulay method. The approach can be understood as the application of the recently developed Alternating Anderson-Jacobi (AAJ) technique \citep{phanish_pask_linear}---an efficient solver for large-scale linear systems in the framework of the classical Jacobi fixed-point iteration---to SCF iterations in electronic structure calculations. Contrary to the conventional wisdom that DIIS generally far outperforms linear mixing, the central idea of the Periodic Pulay method is to employ Pulay extrapolation only once every few SCF iterations, and use linear mixing on all other iterations. We find that this simple generalization not only improves the efficiency of DIIS, but also makes it more robust. In addition, since the majority of electronic structure codes in current use (e.g.,  \cite{kresse1996efficient,Gonze_ABINIT_1,Quantum_Espresso_1,castro2006octopus,soler2002siesta}) already employ Pulay mixing, the proposed technique can be easily incorporated.

The remainder of this letter is organized as follows. In Section~\ref{sec:method}, we present the  Periodic Pulay method. In Section~\ref{sec:results}, we examine its performance for a wide range of materials systems in the context of DFT. Finally, we provide concluding remarks in Section~\ref{sec:conclusions}.


\section{Periodic Pulay method} \label{sec:method}
The self-consistent field (SCF) method casts the equations for the electronic ground-state as the fixed-point problem
\begin{equation} \label{Eqn:fpp}
\br = \bfg (\br) \,,
\end{equation}
where $\br \in \mathbb{R}^{N \times 1} $ is the electron density, and the nonlinear mapping $\bfg:\mathbb{R}^{N \times 1} \rightarrow \mathbb{R}^{N \times 1}$ is composed of the effective potential evaluation for a given electron density and electron density evaluation for the associated Hamiltonian. The convergence properties of the SCF iteration in the vicinity of the solution are determined by the properties of the Jacobian of the residual function $\bff(\br) = \bfg(\br) - \br$ \citep{lin2013elliptic}. Therefore, a strategy that leads to improved conditioning/solvability of the linear system associated with the Jacobian may also lead to improved convergence of the SCF iteration \cite{phanish_restarted_pulay}. In this context, the effectiveness of the GMRES approach \citep{saad1986gmres} in solving linear systems is closely related to the success of the DIIS  method in accelerating SCF iterations \citep{walker2011anderson, fang2009two, rohwedder2011analysis}. Similarly, well established ideas for accelerating the solution of linear systems through preconditioning have found their counterparts in electronic structure calculations \citep{anglade2008preconditioning, ho1982dielectric, kerker1981efficient,lin2013elliptic}. 

In recent work \citep{phanish_pask_linear}, a new solver for large-scale linear systems of equations has been developed in which Anderson extrapolation is performed at periodic intervals within the classical Jacobi fixed-point iteration. On one hand, under-relaxed Jacobi iterations are well known to rapidly damp higher-frequency components of the residual \citep{book_NAPDE_Lui}. On the other hand, periodic application of Anderson extrapolation has the effect of damping lower-frequency components. Therefore, the simultaneous application of these two methods stands to efficiently reduce the overall norm of the residual, thus leading to the success of the so called Alternating Anderson-Jacobi (AAJ) method \citep{phanish_pask_linear}. In particular, AAJ has been found to significantly outperform both GMRES and Anderson-accelerated Jacobi methods. This provides the motivation for the Periodic Pulay mixing scheme proposed here, which can be viewed as the extension of the AAJ method to SCF fixed-point iterations in electronic structure calculations. 

In the Periodic Pulay method, Eqn.~\ref{Eqn:fpp} is solved using the following fixed-point iteration:
\begin{equation}
\br_{i+1} = \br_i + \mathbf{C}_i \bff_i \,, 
\end{equation}
where the subscript $i$ denotes the iteration number, $\bff_i= \bff(\br_i)$, and the matrix 
\begin{equation} \label{Eqn:Bk}
\mathbf{C}_i = \begin{cases} 
                   \alpha \mathbf{I} & \mbox{if } (i+1)/k \not\in \mathbb{N} \,, \quad \text{(Linear mixing)}\\
                   \alpha \mathbf{I} - (\mathbf{R}_i + \alpha \mathbf{F}_i)(\mathbf{F}_i^{T}\mathbf{F}_i)^{-1} \mathbf{F}_i^{T} & \mbox{if } (i+1)/k \in \mathbb{N} \,. \quad \text{(Pulay mixing)}
                   \end{cases}
\end{equation} 
In the above expression, $\mathbf{R}_i$ and $\mathbf{F}_i$ denote the iterate and residual histories:
\begin{eqnarray}
\label{Eqn:Ri}
\mathbf{R}_i & = & \begin{bmatrix} \Delta \br_{i-n+1}, & \Delta \br_{i-n+2}, & \ldots, & \Delta \br_i \end{bmatrix}  \in \mathbb{R}^{N\times n} \,, \\
\label{Eqn:Fi}
\mathbf{F}_i & = & \begin{bmatrix} \Delta \bff_{i-n+1}, & \Delta \bff_{i-n+2}, & \ldots, & \Delta \bff_i \end{bmatrix} \in \mathbb{R}^{N\times n} \,,
\end{eqnarray}
where $\Delta \br_i = \br_i - \br_{i-1}$ and $\Delta \bff_i = \bff_i - \bff_{i-1}$. In addition, $\alpha$ is the mixing parameter, $n$ is the size of the mixing history, and $k$ is the frequency of Pulay extrapolation. We outline the aforedescribed approach in Algorithm \ref{Algo:pPulay}, wherein $tol$ specifies the residual convergence criterion. 

\bigskip
\begin{algorithm}[H] \label{Algo:pPulay}
{\bf Input:} $\br_0$, $\alpha$, $n$, $k$, and $tol$\\ 
\Repeat (\,$i=0,1,2 \ldots $){$\normp{\bff_i} < tol$}
{ 
$\bff_i = \bfg(\br_i) - \br_i$ \\
\eIf{$(i+1)/k \in \mathbb{N}$}
      { $\br_{i+1} =  \br_i +  \alpha \bff_i - (\mathbf{R}_i + \alpha \mathbf{F}_i)(\mathbf{F}_i^{T}\mathbf{F}_i)^{-1} \mathbf{F}_i^{T} \mathbf{f}_i$ }   
      {$\br_{i+1}=\br_i + \alpha \bff_i$}  } 
{\bf Output: $\br_i$}
\caption{Periodic Pulay method}
\end{algorithm}
\bigskip

The Periodic Pulay method can be considered as a generalization of both the classical Pulay and linear mixing schemes. Specifically, the classical Pulay scheme is recovered for frequency of extrapolation $k=1$, while the linear mixing scheme is recovered as $k \rightarrow \infty$. In principle, the parameter $k$ is arbitrary and independent of the mixing history size $n$. However, it is worthwhile to narrow the parameter space for $k$, particularly from the perspective of practical calculations. For this purpose, we note that larger values of $k$ typically lead to more stable but slowly converging SCF iterations (a consequence of increased linear mixing), while smaller values of $k$ tend to provide less damping of higher-frequency error components and correspondingly slower overall convergence as well. In practice, a balance between these limits is preferable and we have found that it is usually counterproductive to set $k > n / 2$ when $n$ is even, and $k > (n+1)/ 2$ when $n$ is odd; and similarly counterproductive to set $k<2$. 

The effectiveness of alternating Pulay and linear-mixing iterations has been recognized before. 
In particular, the Guaranteed Reduction Pulay (GR-Pulay) scheme \citep{GR_Pulay_Bowler,castro2006octopus} alternates classical Pulay and linear mixing in successive SCF iterations with mixing parameter $\alpha=1$. As such, GR-Pulay can be understood as a special case of Periodic Pulay with $\alpha=1$ and $k=2$. However, in more difficult cases, e.g., highly inhomogeneous and/or metallic systems at low temperature, fixing $\alpha=1$ can degrade performance or lead to SCF divergence. The flexibility of reducing $\alpha$ in such cases is thus important to retain. The flexibility to vary $k$ as well---within prescribed limits---can also be advantageous, as we show below. 

While the mixing parameters in the Pulay and linear extrapolations are in general distinct, we use the same mixing parameter for both in the present work. Indeed, the approach can be further generalized by employing different values for the two parameters. However, we have found that such a strategy does not yield significant gains in practice, and therefore refrain from introducing this additional parameter here. {Also, it is worth noting that the Periodic Pulay method does not rely on any heuristics based on the variation of the total energy or residual during the SCF iteration. And yet, as demonstrated in the next section, we find the method to be both robust and efficient for the full range of systems considered, even when classical Pulay fails to converge.} Finally, though we have described the Periodic Pulay scheme in terms of density mixing, it is identically applicable to potential mixing, the expressions for which can be obtained by replacing the electron density appearing in Eqns. \ref{Eqn:fpp}--\ref{Eqn:Fi} with the relevant potential.


\section{Results and discussion} \label{sec:results}
In this section, we verify the robustness and efficiency of the Periodic Pulay method for accelerating the self-consistent field iteration in density functional theory calculations. For this purpose, we implement the Periodic Pulay scheme in the SIESTA code \citep{soler2002siesta,artacho2008siesta}, wherein mixing is performed on the density matrix. To demonstrate the effectiveness of the method across diverse physical systems, we consider the systems listed in Table~\ref{table:materials_systems}. The last four are standard test cases in the SIESTA package. The systems encompass insulating and metallic systems; clusters, surfaces, and bulk phases; and systems with magnetic properties. In addition, many of the systems possess large inhomogeneities in the electron density, which can make SCF convergence more challenging \cite{lin2013elliptic}.

\begin{table}[h!]
\centering
\begin{tabular}{ | c | c |}
\hline
System & Description \\\hline\hline
benzene &  Benzene molecule ($12$ atoms)\\\hline
32-$\text{H}_2\text{O}$ & $32$ molecules of water ($96$ atoms) \\\hline 
Pd-bulk & $3 \times 3 \times 3$ FCC bulk Palladium ($108$ atoms)
\\ & $\Gamma$-point calculation. \\\hline
$\text{Pt}_{13}$ & Octahedral platinum cluster ($13$ atoms)\\\hline
$\text{Fe}_3$-noncollinear & Iron cluster with non-collinear spin ($3$ atoms)\\\hline
Fe-bulk & Bulk BCC iron (1 atom). Brillouin zone \\ & integration with $63$ k-points. \\\hline
SiC-slab & Silicon carbide slab with \\ & hydrogen saturated surface ($78$ atoms) \\\hline
ptcda-Au  & Molecules of 3, 4, 9, 10 perylenetetracarboxylic dianhydride \\ & adsorbed onto gold (111) surface ($204$ atoms)
\\\hline
\end{tabular}
    \caption{Materials test cases.}
\label{table:materials_systems}
\end{table}

In all cases, we compare the performance of the Periodic Pulay method with the classical Pulay (DIIS) method, as implemented in SIESTA. Consistent with usual practice, we consider mixing history sizes of $n \in \{3, 4, \ldots, 8 \}$ for both approaches. Within Periodic Pulay, we vary the frequency of Pulay extrapolation $k$ between $2$ and $n/2$ or $(n+1)/2$, depending on whether $n$ is even or odd, respectively. Unless specified otherwise, we use Fermi-Dirac occupation of the electronic states at temperature $100$ K. We use a mixing parameter of $\alpha = 0.05$ for all systems except benzene and 32-$\text{H}_2\text{O}$, for which we use $\alpha=0.25$. These values of $\alpha$ ensure stable and efficient SCF convergence for classical Pulay mixing. We employ spin polarization in all simulations except benzene and 32-$\text{H}_2\text{O}$. For convergence of the SCF iteration, we use tolerances of $10^{-5}$ in the residual of the density matrix ({max norm}) and $10^{-5}$ eV in the ground-state energy. 

\begin{figure}[h!]
\centering
\begin{subfigure}[b]{0.49\textwidth}
\centering
\includegraphics[keepaspectratio=true,width=0.90\textwidth]{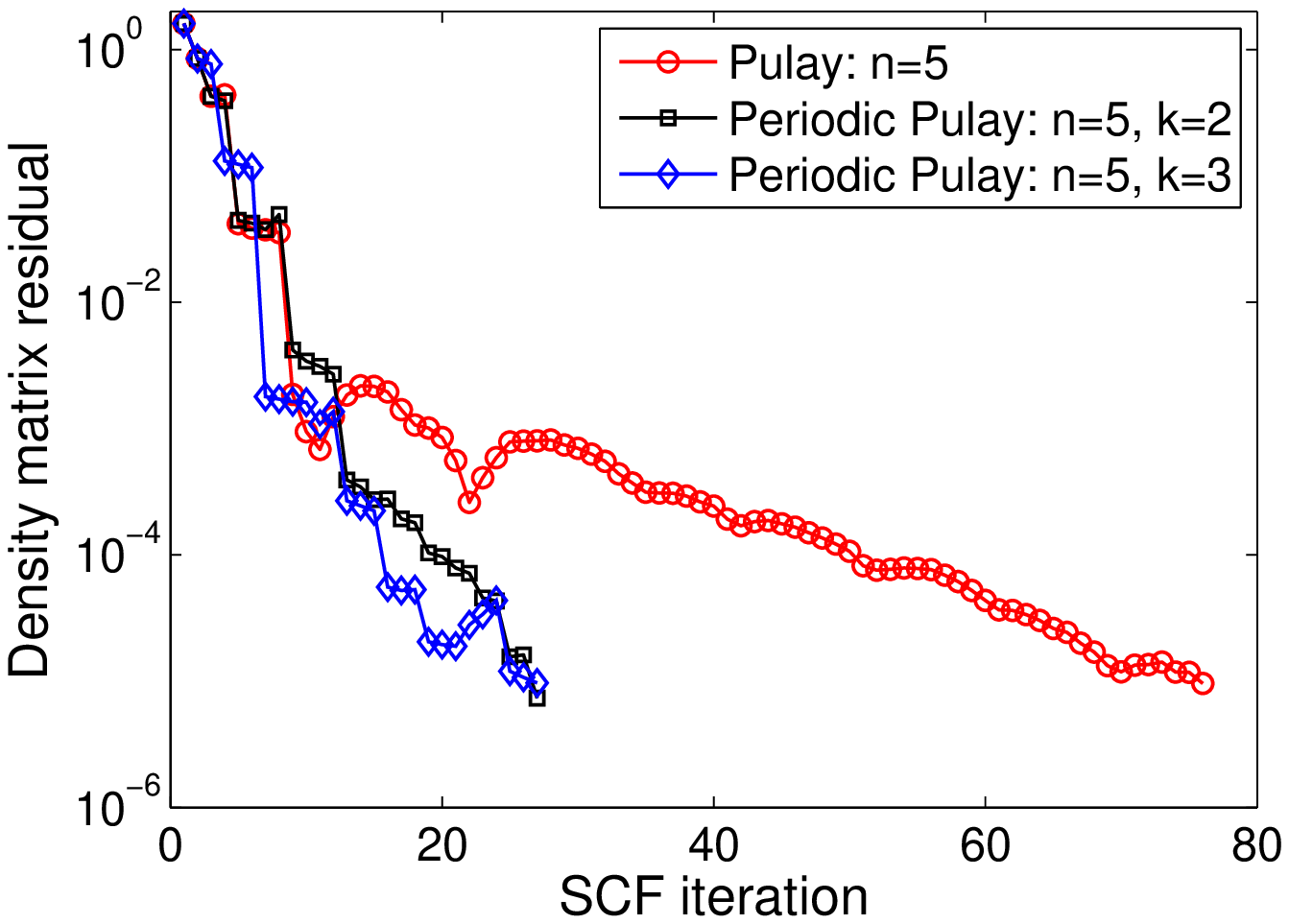}
\caption{Pd-bulk}
\label{subfig:pd_bulk}
\end{subfigure}
\begin{subfigure}[b]{0.49\textwidth}
\centering
\includegraphics[keepaspectratio=true,width=0.90\textwidth]{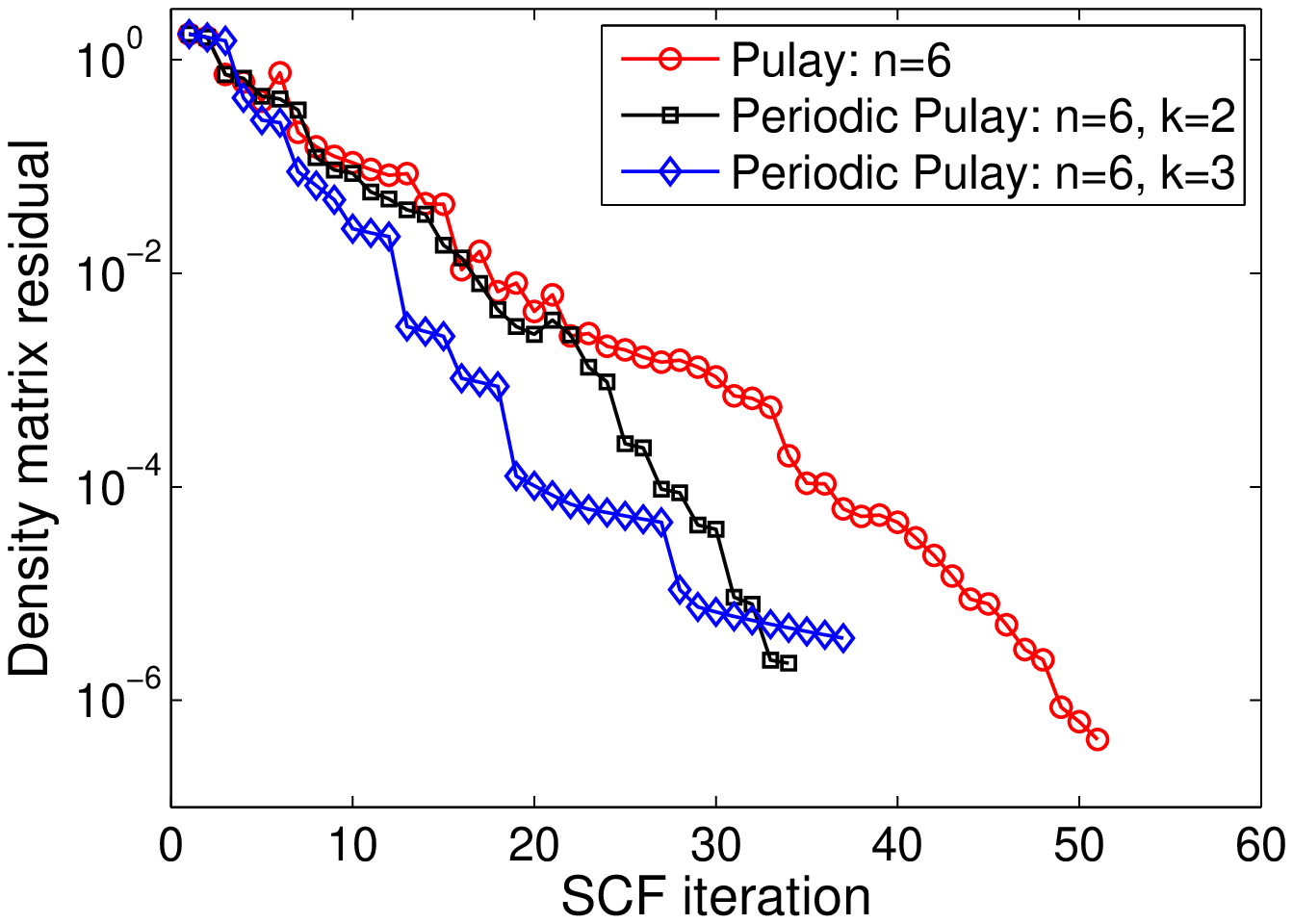}
\caption{SiC-slab}
\label{subfig:sic_slab}
\end{subfigure}
\caption{Comparison of SCF convergence for the Pulay and Periodic Pulay mixing schemes. The mixing parameter $\alpha=0.05$. }
\label{fig:convergence}
\end{figure}

We begin by comparing the convergence properties of the Pulay and Periodic Pulay methods. Specifically, we choose Pd-bulk and SiC-slab systems as representative metallic and insulating cases, for which we use mixing history sizes of $n=5$ and $n=6$, respectively. Additionally, we employ Pulay extrapolation frequencies of $k\in \{2,3\}$ in Periodic Pulay. In Fig.~\ref{fig:convergence}, we present the convergence of the density matrix residual during the SCF iteration. We observe that Periodic Pulay converges rapidly for both values of $k$, and noticeably outperforms DIIS in the process. In fact, for the Pd-bulk system, Periodic Pulay requires a factor of $3$ fewer SCF iterations to bring the residual to the target accuracy of $10^{-5}$. The figures are suggestive of the manner in which convergence tends to be achieved in Periodic Pulay: the residual decreases somewhat slowly over the linear mixing steps (as expected), while the periodic Pulay extrapolation drives the residual down substantially. This is consistent with the behavior observed for the Alternating Anderson-Jacobi (AAJ) method in the context of the linear Jacobi fixed-point iteration \citep{phanish_pask_linear}. 

The two parameters common to the Pulay and Periodic Pulay methods are the mixing history size $n$ and the mixing parameter $\alpha$. These are typically optimized by the user for the physical system of interest. We first consider the performance of the Pulay and Periodic Pulay methods for fixed $n$ and varying $\alpha$. We again choose Pd-bulk and SiC-slab as representative systems, with aforementioned $n$ and $k$ values. We begin by performing an initial traversal of the parameter space for $\alpha$, during which we find that Periodic Pulay not only requires fewer SCF iterations than Pulay in nearly all cases, but is also less sensitive to variations in $\alpha$. In fact, for the Pd-bulk system, we find that Pulay is unable to converge in $250$ iterations with $\alpha = 0.15$, while Periodic Pulay converges in $69$ and $82$ iterations for $k = 2$ and $k = 3$, respectively. We next consider a range of $\alpha$ values including those most efficient for Pulay, i.e., $\alpha \in \{0.01, 0.02, \ldots, 0.10 \}$ and $\alpha \in \{0.10, 0.11, \ldots, 0.20 \}$ for the Pd-bulk and SiC-slab systems, respectively. For these $\alpha$, we present the mean, standard deviation, maximum, and minimum of the number of SCF iterations required to reach convergence in Table 2. We observe that Periodic Pulay has smaller values of both mean and standard deviation, and that its best is always better and worst is never worse than classical Pulay. Hence, we find that on average Periodic Pulay consistently outperforms classical Pulay for fixed n, even for $\alpha$ tuned for classical Pulay.

\begin{table}[h!]
\begin{center}
\begin{tabular}{|c | c c  c c | c c  c c |}
\hline  
    \multirow{2}{*}{System}  & \multicolumn{4}{c|}{Pulay } & \multicolumn{4}{c|}{Periodic Pulay} \\   
                             & $\mu$   & $\sigma$  &  Max & Min & $\mu$   & $\sigma$  & Max & Min  \\\hline
Pd-bulk      & $56$   & $18$    &  $77$ & $27$    &  $36$  & $10$ &  $52$ & $19$ \\                             
SiC-slab  &  $26$  & $4$  &  $35$ & $22$    & $22$   & $2$ &  $26$ & $19$ \\
\hline
\end{tabular}
    \caption{Convergence statistics for Pulay and Periodic Pulay mixing schemes with mixing parameter $\alpha \in \{0.01,0.02,...,0.10\}$ and $\alpha \in \{0.10,0.11,...,0.20\}$ for Pd-bulk and SiC-slab systems, respectively. Mixing history sizes for Pd-bulk and SiC-slab are n = 5 and n = 6, respectively. Pulay extrapolation frequency in Periodic Pulay is $k \in \{2, 3\}$. The mean, standard deviation, maximum, and minimum of the number of SCF iterations required to reach convergence are denoted by $\mu$, $\sigma$, Max, and Min, respectively.}
\label{table:alpha_variation_optim_scan}
\end{center}  
\end{table}

Next, we compare the performance of the Pulay and Periodic Pulay techniques for all systems in Table~\ref{table:materials_systems} for fixed mixing parameter $\alpha$ and varying mixing history size $n \in \{3, 4, \ldots, 8 \}$. Within Periodic Pulay, we consider Pulay extrapolation frequencies $k \in \{2, 3, \ldots, n/2 \, \text{or}\, (n+1)/2\}$. In Table~\ref{table:statistics}, we present the mean, standard deviation, maximum, and minimum of the number of SCF iterations required to reach convergence. We observe that Periodic Pulay has smaller values of both mean and standard deviation for all systems studied. Moreover, the best performance of Periodic Pulay is significantly better than Pulay, while the worst performance of Periodic Pulay is never worse than Pulay. The gains in efficiency are particularly pronounced for metallic systems. We thus find that, while for isolated $\{n,k\}$ combinations Periodic Pulay can occasionally require more iterations, 
on average Periodic Pulay is significantly more efficient and robust than classical Pulay with respect to mixing history size, for all systems studied.

\begin{table}[h!]
\centering
\begin{tabular}{|c | c c  c c | c c  c c |}
\hline  
    \multirow{2}{*}{System}  & \multicolumn{4}{c|}{Pulay } & \multicolumn{4}{c|}{Periodic Pulay}\\   
                             & $\mu$   & $\sigma$  &  Max & Min & $\mu$   & $\sigma$  & Max & Min\\\hline
 benzene                  &  $12$  & $3$  &  $14$ & $9$    & $11$   & $2$ &  $13$ & $9$\\
 32-$\text{H}_2\text{O}$                   &  $19$  & $3$        & $23$  & $16$ & $17$   & $2$ &  $21$ & $16$\\
 Pd-bulk       & $60$   & $40$    &  $125$ & $24$    &  $35$  & $20$ &  $85$ & $22$\\   
 $\text{Pt}_{13}$                   &  $98$  &  $33$    &  $157$ & $74$   &  $76$  & $16$ &  $113$ & $62$\\
$\text{Fe}_3$-noncollinear                &  $112$  &  $38$       &  $173$ & $70$ &  $77$  &  $24$ &  $100$ & $65$\\
Fe-bulk                &  $72$  &  $31$   &  $122$ & $43$    &  $29$  &  $5$ &  $36$ & $23$\\
SiC-slab   &  $51$  & $10$  &  $64$ & $41$    & $31$   & $3$ &  $37$ & $26$\\
 ptcda-Au                   &  $89$  & $11$   &  $103$ & $72$     & $75$   & $8$ &  $97$ & $65$\\
\hline
\end{tabular}
    \caption{Convergence statistics for Pulay and Periodic Pulay mixing schemes for mixing history size $n \in \{3,4,...,8\}$ and Pulay extrapolation frequency $k \in \{2,3,...,n/2 \,\, \text{or} \,\, (n + 1)/2\}$. Mixing parameter $\alpha = 0.25$ for the first two systems, and $\alpha = 0.05$ for the remaining systems. The mean, standard deviation, maximum, and minimum of the number of SCF iterations required to reach convergence are denoted by $\mu$, $\sigma$, Max, and Min, respectively. }
\label{table:statistics}
\end{table}

Metallic systems are well known to face SCF convergence issues as the electronic temperature is lowered, particularly close to absolute zero \citep{lin2013elliptic}. In view of this, we investigate the performance of the Pulay and Periodic Pulay approaches as the electronic temperature is reduced. In Table~\ref{table:temp_variation}, we present the number of SCF iterations required to achieve convergence at different electronic temperatures for the Fe-bulk system.  On one hand, we observe a rapid increase in the number of iterations in Pulay's scheme as the temperature decreases. In fact, it fails to converge in $250$ iterations at an electronic temperature of $25$ K. On the other hand, there are relatively minor variations in the number of iterations required by the Periodic Pulay method, with convergence readily achieved even at $25$ K. These results indicate that the benefits of using the Periodic Pulay method can be more dramatic as the electronic temperature is lowered and SCF convergence becomes more difficult. It is worth mentioning that Periodic Pulay also consistently outperforms Pulay at high temperatures. For example, at $1000$ K, Pulay needs $52$ iterations, whereas Periodic Pulay requires $17$ and $38$ iterations for $k=2$ and $k=3$, respectively.

\begin{table}[h!]
\centering
\begin{tabular}{|c | c | c c |}
\hline  
    Temperature (K) & {Pulay } & \multicolumn{2}{c|}{Periodic Pulay}\\   
                               &      & $k = 2$   & $k = 3$  \\\hline
   \textit{$100$}                   &  $78$       & $25$   & $36$\\
\textit{$50$}   &  $164$  & $75$        & $37$\\
 \textit{$25$}                   &  $\star$  & $39$         & $64$\\
\hline
\end{tabular}
    \caption{Number of SCF iterations required by Pulay and Periodic Pulay mixing schemes at different electronic temperatures for Fe-bulk system (see Table \ref{table:materials_systems}). Mixing history size $n = 5$, and * indicates convergence was not reached in $250$ SCF iterations.}
\label{table:temp_variation}
\end{table}

Finally, we note that performance gains similar to those described above have also been observed when Periodic Pulay is employed in the context of potential mixing within ClusterES \citep{banerjee2015spectral}, a recently developed spectral code for isolated systems. Specifically, factors of $2$ to $3$ speed-up over Pulay are regularly obtained for a variety of clusters. These gains tend to be most pronounced in spin polarized DFT calculations of metallic clusters, in particular. 

Overall, we find that the Periodic Pulay method is relatively insensitive to the choice of input parameters compared to Pulay, which suggests it may be particularly well suited as a ``black-box'' mixing scheme requiring minimal user involvement and parameter optimization. Furthermore, we find that Periodic Pulay generally outperforms Pulay's method for any given set of input parameters. Therefore, the Periodic Pulay method may present an attractive alternative for accelerating SCF iterations in practice.


\section{Concluding Remarks} \label{sec:conclusions}
In this work, we have presented the Periodic Pulay method, a generalization of the classical Pulay (DIIS) method for accelerating self-consistent field iterations in electronic structure calculations. In this approach, Pulay extrapolation is performed at periodic intervals rather than at every SCF iteration, with linear mixing carried out at all other iterations. It was shown that the proposed generalization significantly improves the efficiency and robustness of the DIIS approach for the full range of materials systems considered, encompassing insulating and metallic, bulk and cluster, magnetic and nonmagnetic. Since the vast majority of electronic structure codes in current use employ classical Pulay mixing, and the generalization to Periodic Pulay is straightforward, the potential impact of the new methodology on practical calculations stands to be both measurable and immediate. While our initial analysis suggests that the effectiveness of the new methodology stems, in part at least, from the complementary effectiveness of the linear- and Pulay-mixing steps at reducing smaller and larger wavelength components of the residual, a more rigorous, quantitative understanding remains to be established. This is a worthy subject for further research.


\section*{Acknowledgement}
This work was performed, in part, under the auspices of the U.S.~Department of Energy by Lawrence Livermore National Laboratory under Contract DE-AC52-07NA27344. Support for this work was provided through Scientific Discovery through Advanced Computing (SciDAC) program funded by U.S.~Department of Energy, Office of Science, Advanced Scientific Computing Research and Basic Energy Sciences. P.S.~acknowledges the support of the National Science Foundation under Grant Number 1333500. This work was partially carried out while A.S.B was at the University of Minnesota, Minneapolis. A.S.B acknowledges support from the following grants while at Minnesota: AFOSR FA9550-15-1-0207, NSF-PIRE OISE-0967140, ONR N00014-14-1-0714 and the MURI project FA9550-12-1-0458 (administered by AFOSR). The authors would like to thank the Minnesota Supercomputing Institute for making the computing resources used in this work available.

\bibliographystyle{elsarticle-num}
\bibliography{References}
\end{document}

%% file: preamble.tex
\newcommand{\bff}{{\bf f}}
\newcommand{\bfg}{{\bf g}}

\newcommand{\beq}{\begin{equation}}
\newcommand{\eeq}{\end{equation}}
\newcommand{\beqs}{\begin{eqnarray}}
\newcommand{\eeqs}{\end{eqnarray}}
\newcommand{\beql}{\begin{equation} \label}

\let\oldFootnote\footnote
\newcommand\nextToken\relax

\renewcommand\footnote[1]{%
    \oldFootnote{#1}\futurelet\nextToken\isFootnote}

\newcommand\isFootnote{%
    \ifx\footnote\nextToken\textsuperscript{,}\fi}